**Michiel Hazewinkel**     1     CWI
Direct line: +31-20-5924204     POBox 94079
Secretary: +31-20-5924058     1090GB Amsterdam
Fax: +31-20-5924166
E-mail: mich@cwi.nl     original version: 12 February 2004
    revised version:

# Mathematical knowledge management is needed
(Keynote speech at the November 2003 MKM meeting in Edinburg)

by
*Michiel Hazewinkel*
*CWI*
*POBox 94079*
*1090GB Amsterdam*
*The Netherlands*

**Abstract**. In this lecture I discuss some aspects of MKM, Mathematical Knowledge Management, with particuar emphasis on information storage and information retrieval.

**MSCS**: 68P20, 68T30

**Key words and key phrases**: mathematical knowledge, knowledge management, IRIS, information retrieval, information storage, keyphrase assignment, identification cloud, formal mathematics, classification, metadata, MSCS.

### 1. Is there a problem?

The issue at hand is that of information storage and information retrieval as regards mathematics. Or, concentrating on the latter, if one has a mathematical question, can one find out what is known about it (in reasonable time). Many say yes; what with friends (= knowledgeable colleagues), two quite good databases of abstracts (complete with sophisticated search engines), and, lately, the web with a good many search engines and full text search, of course we can. I am of a different opinion and in this lecture I will try to indicate why I think that:

"We don't even know how much we know that we don't know we know".

Mostly the reasons why I think this come from my experiences as editor of the Encyclopaedia of Mathematics, [8], and, consequently, trying out to find what is known in areas where one is not a (super)specialist (meaning a highly specialized scientist).
 Of course this is not unique to mathematics. As far as sheer quantity of published material goes, physics, chemistry, biology, medicine, ...  are far larger. Think of a factor 10 or so. On the other hand, results in mathematics have a long life. As an example, results from around 1850, that were all but forgotten, can be very relevant for questions to be solved now (as I can testify from personal experience; the results in question were some of Kronecker concerning cyclotomic units).

### 2. The size of the problem.

To start with let me give you some descriptive numbers of the size of this matter, the sheer volume of paper (data) involved.
 The total number of pages published in mathematics so far makes up a stack of about 60km height. This is excluding college level textbooks such as those voluminous texts on calculus and analytic geometry beloved by certain publishers. Or, to put it another way, 60 km shelf space. So, taking back to back shelves, seven high, separated by the absolute minimum of 80cm, one would need a room of 100m by 120m, about the size of two football fields.

How much redundancy there is in all that is anybody's guess. At present there not even preliminary ideas how to estimate that.

Here are some more indicators of size and growth.

The universally used and excellent classification scheme MSCS2000 (Mathematical Subject Classification Scheme, version of the year 2000) [1] is a tree with some 5500 leafs. That is a lot of leafs. Still, most of these are large enough to drown a whole (super)specialism in. Here are some examples.

14L05: Formal groups, p-divisible groups. The 570 page monograph [7] contains maybe half of what is necessary to give an adequate treatment of the subject.

16W30: Coalgebras, bialgebras, Hopf algebras. The time is long past that the subject of Hopf algebras could be dealt with reasonably in one monograph; 2500 pages is my estimate.

20D08: Sporadic groups. Besides some small number of infinite series there are just 26 socalled sporadic simple groups. Describing them can be done in a few hundred pages; proving that these are all would take well over 5000 pages. Indeed just dealing with the largest of them, the Fischer-Griess monster, can be the subject of a full length monograph (and in fact such a monograph is being written).

90C27: Combinatorial optimization. Recently, 2003, there appeared a three volume monograph on this subject totalling 1882 pages, [18]. According to the author this represents just those parts of the subject that were interesting to him (private communication).[2]

John Ewing in a recent article in the Mathematical Intelligencer, complaining about the excessive[3] (his words) profits made by commercial publishers, noted that there were some 25000 mathematical articles published in 2001.

In connection with [8], I once calculated in several different ways, that to give an adequate description of the more established parts of mathematics one would need between 120 000 and 200 000 (controlled standardized) key phrases. The first 10 volumes of loc. cit. have about 30000 (not counting inversions and linguistic variations. So to do a good job a four to six fold increase is needed. Work on that is in progress.

In the 1970's Raoul Bott of Harvard had a graduate student he did not really know what to do with. So he set him to counting how many new theorems there appeared a year (using Math

---

[1] To see the scheme go to www.emis.de/ZMATH and click on the second item under the heading 'services' in the left side bar.

[2] These considerations suggest that all of mathematics can be dealt with in some 11 000 monographs; thus reducing the stack of 60km to someting like 550m, assuming that enough competent people can be induced to do the job. There are only some 50 000 mathematicians worldwide (of which much less than 50% are active). This gives maybe a first idea of how much redundancy is involved. However, the idea that one can cover all of mathematics in some 100 volumes, as the publisher Springer once claimed in an advertisement for the series EMS, is of course utter nonsense.

[3] I tend to diasagree. The whole debate sounds to me like the squeeking of a mouse that has seen a particle of cheese slide away while next door a large rat is gnawing contentedly on a big chunk of meat. Many seem to be complaining about the price of scientific literature. At the very least the debate should include the price and the vastly larger profits involved in manifacturing the instruments of scientific research and practice in physics, astronomy, chemistry, biology, and, especially, medicine.

Review). The number was 200 000. These are what their respective authors called theorems; not lemmas, propossitions, scholia, constructions, definitions, ... .

There is a pleasing thing about these various numbers, coming from very different ways of looking at the field of mathematics. They all, within an order of magnitude, fit with one another.

It is a remarkable fact that vast sums, worldwide, are spent on research and very little on making sure that if and when some known result is needed it can be found again.

This leads to the well known and widespread phenomenon of rediscovery. For instance I once read a paper from algebraic topology which contained a substantial amount of finite group theory which was needed; later in the review in Mathematical Reviews the reviewer complimented the author on the fact that he had not only done important work in algebraic topology but also in group theory; still later in turned out that the group theory part had been done long before.

It has been said that for moderately difficult problems it is often less effort to solve them again than to try to find them in the published literature.[4]

A famous case (of rediscovery) occurred in the theory of completely integrable dynamical systems. An important question there is that of commuting differential operators. A lot of work on that had been done when it was discovered that the question had been settled in 1928 by Burchnall and Chaundy, see [3, 5]. For another tool in the field of completely integrable systems, that of what are now called 'Akhiezer functions', a similar story is true.[5] For a while, indeed, it was something of a cult thing in integrable system theory to look for and find old forgotten results that were important for the theory.

### 3. How old is the problem?

Actually quite old. Burton wrote in his 'Anatomy of melancholy', first edition published in 1621, when discussing ways of combating that very 'melancholy' (Part 2, Sect. 2, Memb.4; p. 455 in [6]):

"What vast Tomes are extent in Law, Physick, & Divinity, for profit, pleasure, practice, speculation, in verse or prose, &.! Their names alone are subjects of whole Volumes, we have thousands of Authors of all sorts, many great Libraries full well furnished, like so many dishes of meat, served out for several palates; & he is a very block that is affected with none of them."

and, ibid p. 460:

"By this art you may contemplate the variation of the 23 letters, which may be so infinitely varied, that the words contemplated and deduced thence will not be contained within the compass of the firmament; ..."

or, quoting a much more recent author:

"It suffices that a book be possible for it to exist."

<div style="text-align: right">J L Borges, The library of Babel</div>

However, reading these old accounts, one does not get the feeling that the abundance of learning then extant was regarded as a problem. Rather it was regarded with warm feeling, like a treasure house affectionatically viewed with the comfortable sentiment that there will always be enough for all.

Perhaps that was the case because at the time, Burton's time, scientists mainly relied on human memory and human information processing. The human brain is a magnificent information processing device (operating in very different ways from computers). Just how good it is (can be) appears to be largely unknown. Estimates of the memory capability vary wildly from $10^{20}$ bits (von Neumann) to $10^9$ bits (Th K Landauer, 1986). Estimates based on the number of synapses range from $10^{13}$ to $10^{15}$ bits of memory. Clearly, given the astonishing feats of some people with an eidetic (photographic) memory the potential is very high. Personally I am inclined to be very optimistic about the (potential) information processing capabilities of humans and to think that it was/is a mistake to trust too much to computers (and before that to card indexing systems).

---

[4] And, most appropriately, I cannot at the moment find where the algebraic topolgy example came from.

[5] But in this case the results go still further back to the first years aof the 20-th century.

## 4. Why MKM?

MKM stands for 'Mathematical Knowledgement Management'. Whence the theme of this conference. Also the inspiration. Whether it is explicitely realized/acknowledged or not, many national and international initiatives have to do with MKM, or ?KM. As distinguished from M(KM), which is the mathematics of knowledge management, a parsing which we own to Bruno Buchberger, and a subject which has hardly been touched so far, but see [11, 12] and the references cited there, and, perhaps, when in a generous mood, the business of citation analysis.

There is much more to MKM than just mathematical information storage and retrieval. These are by no means the only issues. Others are formalized mathematics, proof checking and transparent proofs, the quality of proofs, indexes and thesauri, mathematical databases, ... . Some of these will briefly be discussed below.

## 5. Tools 1. The MSCS 2000 (Mathematics Subject Classification Scheme, version of 2000).

The field of mathematics is lucky in that it has a very good classification scheme. It is a tree of four levels (counting the root) consisting of the root (all of mathematics), specialisms within mathematics (such as '54: General topology), subspecialisms (such as 35L: Partial differential equations of hyperbolic type) and subsubspecialisms (such as 20C30: Representations of finite symmetric groups). There are about 5500 of these finest classification numbers (leaves).

Perhaps inevitably the thing is organized as a tree in imitation of evolutionary trees. Certainly mathematics and other fields of knowledge evolve, but they do not evolve (like nature) in ever finer subdivisions and more specialized species. Other things happen too. Like two subsubspecialisms merging, or two quite far apart subsubspecialisms turning out to be quite narrowly related, or one subspecialism turning out to be a special case of some subsubspecialism elsewhere. All these things do not happen in nature's evolutionary tree.

A more accurate picture of how the various parts of mathematics interrelate etc. is probably is directed graph and with modern computer technology and computer graphics there is no reason not to display things like that.

Still trees[6] are traditional for classification schemes and thesauri and an awful lot of 'see also' links do help a good deal.

To illlustrate that mathematics is not at all like a tree, here is the part of the MSCS that is relevant for the concept of Hopf algebras

---

[6] Personally I think trees are an inheritance from the printing age: they are just about the most complicated directed graphs that can be conveniently printed.

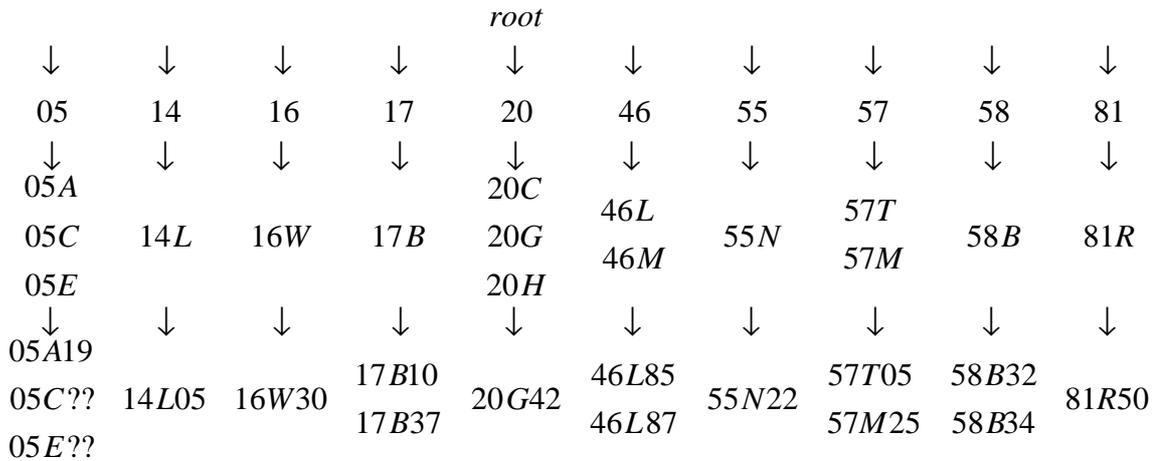

Here are some words on what the diagram above means.

* The exact phrase "Hopf algebras" occurs twice in the MSCS: 16W30 and 57T05. Basically a Hopf agebra is an algebraic structure; more precisely it is an algebra with additional structure. This gives the classification 16W30 (= coalgebras, bialgebras, Hopf algebras and modules on which they act). The reason that there is a classification 57T05: Hopf algebras, is that historically the first examples of Hopf algebras came in the form of the cohomology or homology of suitable spaces and manifolds.

* 05 stands for the subfield ''Combinatorics'. A good many combinatorial identities arise by looking at representations of Hopf algebras in two ways (often universal enveloping algebras, a special kind of Hopf algebras), whence lots of applications of Hopf algebras in subsubfield 05A19 (= Combinatorial identities).

Many combinatorial objects naturally form the basis of a Hopf algebra; for instance planar binary trees. Whence e.g. the role of Hopf algebras in 05C (= Graph theory).

The symmetric functions and several generalizations are best looked at as forming Hopf algebras. Hence the role of Hopf algebras in subfield 05E (Algebraic combinatorics).

* Many important Hopf algebras are commutative. So one would expect a subfield 13??? (13 = Commutative rings and algebras), for commutative algebras with extra structure and a place for Hopf algebras there. But there isn't.

* 14L05 is 'Formal groups, p-divisible groups', a subsubfield of 14 (= Algebraic geometry). Formal groups are a special kind of Hopf algebras which in turn have applications in number theory (11M, 11R, 11S), field theory (12F). Another kind of Hopf algebras is formed by (the coordinate rings of) algebraic groups (14L, 20G, 20H)

* 17B stands for 'Lie algebras', and 17B10 for the algebraic approach to representations of Lie algebras. A representation of a Lie algebras is the same thing as a representation of its universal enveloping algebra which is a special kind of Hopf algebra.

17B37 is ''Quantum groups, quantized enveloping algebras and related deformations'. This kind of defomations is also a special kind of Hopf algebra.

* 20C is 'Representations of groups'. A representation of a group is the same thing as a representation of its group ring which is another special kind of Hopf algebra.

* 20G is the subfield of 'linear algebraic groups' and 20G42 is the subfield of 'quantum groups' which are special kinds of Hopf algebras (deformations of the coordinate rings of the classical groups).

* 46 is the specialism 'Functional analysis' with subfield 46L: 'Selfadjoint operator algebras'. There are well over 300 published papers dealing with Hopf algebras of this kind. Mostly in subsubfields 47L85 (Noncommutative topology), 46L87 (Noncommutative geometry). There are also some in 46M (Methods of category theory in functional analysis)

  * 55N is the subfield 'Homology and cohomology theories' of 55 (= Algebraic topology) and 55N22 is'Bordism, cobordism, formal group laws'. As already remarked 'a formal group is a special kind of Hopf algebra. For that matter the applications of formal group laws in (algebraic) topology are not limited to (co)bordism theories but also take in elliptic cohomology and (extraordinary) K-theories.

  * 57M is 'Low dimensional topology'. There are numerous applications of Hopf algebras (quantum groups) to 57M25 (= Knots and links) and also to other parts of 57M (such as invariants of three manifolds).

  * 58 is the field ''Global analysis, analysis on manifolds', which has as subsubspecialisms 58M32 'Geometry of quantum groups' and 58M34 ''Noncommutative geometry (à la Connes)'.

  * Finally 81 is ''Quantum theory' and 81R is ''Groups and algebras in quantum theory' with the by now quite large subsubfield 81R50: 'Quantum groups and related algebraic methods'. (There is also 81R60: ''Noncommutative geometry' (again).)

Even this is not all. There are some 36 papers in the ZMATH[7] database having to do with Hopf algebras in (theoretical) computer science (semantics, transsition systems, concurrent and distributed computing, accessible data types, ...). There also a small number of papers (8) about Hopf algebras in 'Control and system theory' (area 93), mostly having to do with realization theory (of which the essence seems to lie in the coalgebra part).

As to the total size of the field, a search done on 20 Oct. 2004 in the datbase ZMATH gave 10589 hits for Hopf algebras.[8]

**6. The price of knowledge.**

Suppose you have all 200 or so volumes of the journal ''Theoretical Computer Sciene' standing on your shelves, nicely bound in blue leather. And suppose you need to find out something about, say, 'the use of deontic logic in computer science'. Then there is a bit of a difficulty. The same applies if you have the same problem and are in a large library with, say, the books nicely arranges alphabetically according to first author.

  What are needed are metadata: classifications, key words and key phrases, and all kind of other data that can help a scientist finding what he is looking for.

  This is well known and is why libraries spent a lot of time on 'disclosure', assigning suitable metadata to documents say according to the scheme used by the American Library of Congres or the new emerging standard called the 'Dublin Core'.

What is far less known is how expensive this (totally insufficient) business is. In spite of todays horrible prices for scientific books it is still true that attaching metadata to it (in salary of personel and other expenses (overhead) is more than the price of the book[9].

  It is also why books have indexes[10] and why occasionally an enlightened editor publishes an index of a large number of volumes of his journal.

  It is probably true that the most useful metadata for IRIS (Information Retrieval and

---

[7] Zentralblatt für Mathematik und Grenzgebiete.

[8] Technically , the search was for 'Hopf algebras' or 'bialgebras' (which are very nearly Hopf algebras), or 'quantum groups' or 'quantum algebras' (which are special kinds of Hopf algebras). Group rings and enveloping algebras (which are also special kinds of Hopf algebras were not included.

[9] Private communication.

[10] Mostly very bad and incomplete ones.

Information Storage[11]) purposes is a combination of classifications and standardized key phrases. There is a good classification scheme, but it is to coarse for adequate metadata descriptions of articles and books. On the other hand adding another layer would make it unwieldy and not useful anymore. There remains the idea of using key phrases extensively. There are difficulties with that, there is no standardized keyphrase lists for all of mathematics, though first attempts towards creating such have been made see [8, 9, 10] and others, and of the two big (and very useful) databases, one, MathReview, does not have a key phrase field and the other, ZMATH, only has a free (= wild, i.e. uncontrolled) key phrase field.

Finally, as will be argued below, 'full text search has its own special difficulties.

## 7. Tools 2. Full text search.

A remark that I have often heard when the various problems concerning automatic assignment of keyphrases were discussed goes something like the following: "Why bother with this kind of metadata, now that we (nearly) have full text search capability".

There are quite a number of counter arguments to this; some well known, some not.

In the first place to quote from [14], page 2, "Natural language is easy for people and hard for machines;...". Part of which is that the number of linguistic variations for a multiword keyphrase tends to be very large. Add to that that in science diffferent words may have the same meaning, that the names of the same theorems and concepts can vary between specialisms and geographical localities, ambiguity matters, ... , and one has a major collection of problems to handle.

In the second place it is very well possible that there is a key phrase or concept or ... that is a very good descriptor of a given piece of text but which is simply not present. This is a problem when (full text) searching for that phrase; it is also a problem in the automatic assignment of keyphrases. For both problems I believe the idea of identification clouds can help, provided it is suitably further developed. See [11, 12] for an extensive discussion of identification clouds and [13] for a description of some problems that still need to be solved.

The idea of identification clouds can also do something about what has been called the 'key phrase barrier', [17].

A third reason why full text search is not the end of the story is the story of ends. The fact is that 'end' is a technical concept in mathematics. Actually it is three related but slightly different concepts in function theory (MSCS classification 30 ), group theory (MSCS classifications 20 and 22) and general topology (MSCS classification 54). Suppose now a researcher needs to find out something about the technical concept 'end'. Then searching the full text he will pick up an awful lot of extraneous totally irrelevant material simply because 'end' occurs often in its linguistic sense as in at the end of the chapter; end of a book; end of a paper; the article ends with; end of the 20-th century; the beginning and end of, end of the proof, end of the discussion, .... . Also there are dozens and dozens phrases in these and other parts of mathematics which have 'end' in them. Things would be entirely different if the word 'end' occurred in a dedicated keyphrase field (to be filled in from a list of controlled keyphrases or a controlled enriched weak thesaurus).

There are quite a few such words. Some more examples are 'sort', 'form' (with several distinct mathematical meanings), 'net' (which has at least five different mathematical meanings), 'current', 'triple' (at least three meanings), 'frame', 'picture'. A more word example is 'transparent proof'. There is a technical notion called that due to L Babai a.o., see [2], but a search on 'transparent proof' in ZMATH practically only yield articles in which the phrase occrs

---

[11] Iris is also the name of a Greek goddess. One of her duties was to bring a ewer of water from the river Styx to render unconscious for a year any god wh perjured him/her self; a treatment that should be extended to scientists who falsify or plagiarize. By a not inappropriate quirk of nomenclature 'Iris" is also the name of the electronic submission system of the Dutch organization for funding scientific research.

in its mundane meaning of a lucid, clear proof.

## 8. Tools 3. W4, the wonderful world wide web?

There is also a wide spread delusion that if you need to find anything you can find on the world wide web. There are amny reasons why that is not the case, at least not reliable information.

First there is enormous contamination. Many things on the web are simply not true and also many are deliberate lies and falsifications.

There is also a second source of contamination. There are all kind of oufits that sell services that make their clients websites come up high in the ranking of the various search engines.

Thirdly, according to the study [15, 16] even the best of the search engines, Northern Light, finds only some 16% of the relevant material.

As far as contamination goes things become enormously much better if one concentrates only on science sites. This can be done by going to SCIRUS <www.scirus.com> (which is free).

## 9. Formalized mathematics.

The last topic I would like to touch on is that of the reliability of mathematical knowledge. There is no doubt that standards of mathematical rigour, of mathematical proof have changed in the direction of becoming stricter. There is also no doubt that somewhere in the 5000 pages or so that it takes to prove the classification of the finite simple groups, notably the sporadic groups, there are errors; but equally there is no doubt that the theorem is all right and that the errors can be repaired. Until 1993 the situation was very different with the proof of the four colour theorem by Appell and Haken in 1977, [1]. That was and is an extraordinary fragile proof not least because there was (and is) no proof that the thing can in fact be handled by checking a finite number of unavoidable configurations, still less a bound for them. Things changed in 1993, see [19].

Thus there is quite a difference between 'robust proofs' and ''fragile proofs' which begs to be examined. Possibly the tools of formalized mathematics and/or the idea of 'transparent proofs', [2], can play a role here.

There is also a major problem here. Composing a formal proof of even something quite simple is a substantial amount of work. And thus keeping up with some 300 000 new theorems a year is (at the moment) hopeless.

## 10. What can be done?

Well, rather a lot. There is a substantial amount of activity. To get some idea, see the proceedings of the first MKM meeting in Linz, [4], the lectures at this conference, and the proposal MKMnet (IST-2001-37057).

In particular here are quite a few aspects I have not touched on above: languages for representing and displaying (formal) mathematics, repositories of formal mathematics, tools for organizing (formal) mathematics, proof checkers and theorem provers (automatic reasoning systems), computer assisted mathematics (computer algebra systems, more specialized systems), ... . For a survey of much of what exists in this area see deliverable D4.1 of MKMnet (IST-2001-37057).